\documentclass{article} 
\usepackage{iclr2026_conference,times}

\usepackage{amsmath,amsfonts,bm}

\def\eqref#1{equation~\ref{#1}}

\def\1{\bm{1}}

\DeclareMathAlphabet{\mathsfit}{\encodingdefault}{\sfdefault}{m}{sl}
\SetMathAlphabet{\mathsfit}{bold}{\encodingdefault}{\sfdefault}{bx}{n}

\DeclareMathOperator*{\argmax}{arg\,max}

\usepackage{cite}
\usepackage{hyperref}
\usepackage{url}
\usepackage{paralist}
\usepackage{algorithm}
\usepackage{algpseudocode}
\usepackage{amsmath}
\usepackage{amsfonts}
\usepackage{times}
\usepackage{caption}
\usepackage{multirow}
\usepackage{subcaption}
\usepackage{booktabs}
\usepackage{graphicx}
\usepackage{xspace}
\usepackage{enumitem}
\usepackage{tabularx}

\usepackage{lipsum}
\usepackage{indentfirst}
\usepackage{dsfont}
\usepackage{relsize}
\usepackage{svg}
\usepackage{amsthm}
\usepackage{caption}
\usepackage{listings}
\usepackage{threeparttable}
\usepackage{array}
\usepackage{float}
\usepackage{ragged2e}
\usepackage{xcolor}
\usepackage{amssymb}
\usepackage{tikz}
\usepackage{rotating}
\usepackage{bbm}
\usepackage{xurl}
\usepackage[most]{tcolorbox}
\tcbuselibrary{theorems}
\usepackage{cleveref}

\setlist[itemize]{leftmargin=0.1in, itemsep=0em}
\newtcbtheorem[]{exmp}{Prompt}%
{fonttitle=\bfseries, left=.02in, right=.02in,bottom=.02in, top=.02in}{exmp}
\newtcbtheorem[]{example}{Reasoning Trace}%
{fonttitle=\bfseries, left=.02in, right=.02in,bottom=.02in, top=.02in}{exmp}

\newcommand{\eg}{\textit{e.g.},\xspace}

\definecolor{darkblue}{HTML}{00008B}
\newcommand{\rebuttal}[1]{\textcolor{black}{#1}}

\newcommand{\method}{ParallelResearch\xspace}
\let\cite\citep

\newcommand{\ablation}{\method (-AP, -RO)\xspace}

\title{Efficient Tree-Structured Deep Research\\ with Adaptive  Resource Allocation}

\author{
Lunyiu Nie$^1$\thanks{Work done during internship at Adobe Research.}~, Nedim Lipka$^2$, Ryan A. Rossi$^2$, Swarat Chaudhuri$^1$\\
$^1$The University of Texas at Austin\ \ \ $^2$Adobe Research\\
\url{lynie@utexas.edu}
}

\iclrfinalcopy 
\begin{document}

\maketitle

\begin{abstract}
Deep research agents, which synthesize information across diverse sources, are significantly constrained by the sequential nature of reasoning. This bottleneck results in high latency, poor runtime adaptability, and inefficient resource allocation, making today's deep research systems impractical for interactive applications. 
To overcome this, we introduce \textbf{\method}, a novel framework for efficient deep research that transforms sequential processing into parallel, runtime orchestration by dynamically decomposing complex queries into tree-structured sub-tasks.
Our core contributions are threefold: \textbf{(1)} an \textbf{adaptive planner} that dynamically allocates computational resources based on query complexity; \textbf{(2)} a \textbf{runtime orchestration layer} that prunes redundant paths to reallocate resources and enables speculative execution; and \textbf{(3)} a \textbf{fully-asynchronous execution infrastructure} that enables concurrency across both research breadth and depth. Experiments on two benchmarks show up to 5$\times$ speedups with comparable final report quality, and consistent quality improvements with the same time budgets.
\end{abstract}

\section{Introduction}
Deep research systems, which can generate long-form responses by synthesizing information from
diverse resources given an open-ended query, have become increasingly powerful with the advances in LLM-based agents.
While the state-of-the-art systems can create high-quality reports for deep research tasks, including literature review \cite{haman2025fake} and policy analysis \cite{gambrell2025ai}, their efficiency is still limited: such systems often take tens of minutes to respond. This latency can break users' cognitive flow  \cite{iqbal2007disruption}, incur high context-switching costs \cite{mark2008cost}, and degrade overall experience. 

Much of this inefficiency stems from poor runtime orchestration. Existing systems typically rely on sequential planning and reasoning \cite{xu2025comprehensive}, leading to unnecessary latency when opportunities for parallel branching and speculative exploration are under-utilized. These systems typically plan ahead of research. However, the value of subqueries or deeper investigation often becomes clear only after partial research progress, but current systems rarely prune low-value paths or reallocate resources at runtime.

To address these challenges, we propose \textbf{\method}, a framework designed for efficient deep research that integrates adaptive planning and runtime orchestration on a fully-asynchronous execution infrastructure. \method treats deep research as a dynamic, tree-structured traversal, where a complex query is decomposed into concurrent subqueries that dynamically populate the research tree. The objective is to maximize response quality under a time budget by adjusting the tree structure and reallocating resources across promising research paths.

At the core of \method is an adaptive planner that weighs the expected information gains \emph{before} and \emph{after} each research step, expanding breadth and deepening paths only when the information gain justifies the overheads. This allows \method to flexibly allocate resources depending on nature, scope, and complexity of each query. 

\begin{figure*}[t]
    \centering
    \includegraphics[width=\textwidth]{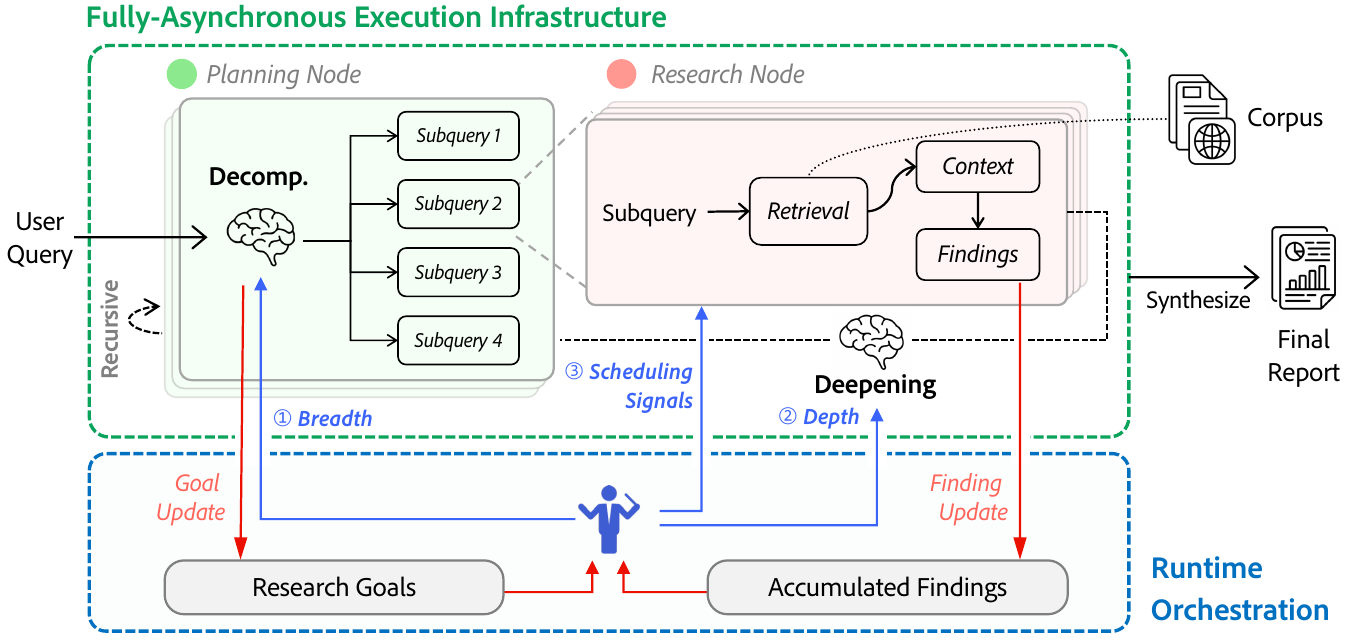}
    \caption{Overview of \method: the Planning Nodes adaptively decompose queries into parallel subqueries executed by Research Nodes for findings, which may recursively trigger deeper planning. The \textbf{adaptive planner} expands (1) \emph{breadth} to explore \underline{prior-research} and regulates (2) \emph{depth} to pursue promising paths \underline{post-research}. The \textbf{runtime orchestration layer} monitors progress and reallocates resources through (3) \emph{scheduling signals} \underline{mid-research}. A \textbf{fully-asynchronous execution infrastructure} enables flexible concurrency across both breadth and depth.     }
    \label{fig:architecture}
\end{figure*}

However, planning alone is insufficient. Since research is inherently iterative and non-linear, newly emerged evidence \emph{mid-research} may reshape priorities. \method incorporates a runtime orchestration layer that monitors ongoing research findings, evaluates them against goal satisfaction and quality rubrics, and makes live adjustments. This mechanism allows the system to terminate low-value or redundant branches early and reallocate computational resources toward more promising paths. More importantly, it allows \emph{speculative execution} by launching child tasks before parent-level planning decisions are finalized, thereby reducing latency and accelerating throughput.

This tight feedback loop between planning and research execution produces a highly dynamic research tree that evolves at runtime. To handle this, \method employs a fully-asynchronous execution infrastructure to schedule all the planning, research, and orchestration tasks in a unified pool with thread-safe state management. This enables concurrent exploration of multiple research paths and allows non-blocking orchestration to adapt the tree structure.

To evaluate the effectiveness of \method, we conduct experiments on two recent deep research benchmarks, DeepResearchGym \cite{coelho2025deepresearchgym} and DeepResearch Bench \cite{alzubi2025open}. Compared to the baseline, \method can consistently improve research throughput, producing research reports of the same quality with up to 5$\times$ speed-ups, or even higher quality within the same time. 

The key contributions of this work are:
\begin{compactitem}
    \item A formal formalization of deep research tasks as a tree-structured optimization problem.
    \item An adaptive research planner for dynamic, context-aware decisions on task branching and deepening.
    \item A runtime orchestration layer for task monitoring, speculative execution, and resource reallocation.
    \item A fully asynchronous execution infrastructure enabling concurrent planning, research, and orchestration across breadth and depth of the evolving tree.
    \item A comprehensive empirical evaluation on three model families across two benchmarks, demonstrating \method's significant improvements in terms of research throughput, quality, and efficiency.
\end{compactitem}

\section{Related Works}
\subsection{Deep Research Agents}
Building on earlier tool-use frameworks like WebGPT \cite{nakano2021webgpt} and ReAct \cite{yao2023react}, recent deep research agents -- including GPT-Researcher \cite{gptresearcher}, Open Deep Search \cite{alzubi2025open}, and LangChain’s Open Deep Research \cite{langchain} -- decompose complex queries into tool-augmented subtasks. To standardize evaluation, emerging benchmarks like DeepResearchGym \cite{coelho2025deepresearchgym} and DeepResearch Bench \cite{du2025deepresearch} introduce LLM-as-a-judge protocols tailored for complex research questions. 

Despite these advances, current systems typically rely on fixed, pre-specified parameters for controlling the research structure. Their orchestration strategies are dominated by sequential execution or coarse-grained parallelism \cite{xu2025comprehensive}, which limits adaptability. As a result, when information quality shifts during execution, these systems either waste compute or incur unnecessary latency. \method addresses these limitations by introducing a runtime orchestration layer that couples fully-asynchronous execution across both depth and breadth. Unlike prior static approaches, it adaptively expands or prunes subqueries in real time based on intermediate evidence, enabling more efficient and responsive deep research.

\subsection{Parallel and Speculative Reasoning}
Token- and action-level acceleration methods such as speculative decoding \cite{leviathan2023speculative, miao2023specinfer, cai2024medusa} and speculative reasoning for fast inference \cite{pan2025specreason, yang2025speculative} reduce latency via draft-and-verify or multi-token prediction. At the reasoning level, Dynamic Parallel Tree Search \cite{ding2025dynamic} accelerates Tree-of-Thoughts by expanding and pruning nodes in parallel, while ParaThinker \cite{wen2025parathinker} and Parallel-R1 \cite{zheng2025parallel} instill native parallel reasoning. These improve efficiency and accuracy but still rely on static branching.
Inspired by these works, \method advances parallelism to the workflow level: it not only reallocates compute and prunes branches dynamically, but also supports speculative execution—allowing branches to expand without delay and later discarding them if evidence shows they are unnecessary.

\subsection{Agent Workflow Orchestration}
Recent work compiles high-level goals into executable agent graphs via MCTS-guided code search \cite{zhang2024aflow}, evolutionary populations of heterogeneous workflows \cite{niu2025flow}, and modular activity-on-vertex graphs \cite{zhang2025evoflow}. 
These systems mainly optimize \emph{offline} and then execute largely fixed graphs, and their runtime control over partially executed graphs remains limited. 
Production frameworks such as AutoGen \cite{autogen}, LangGraph \cite{langgraph}, DSPy \cite{dspy}, and OpenAI Swarm \cite{openai_swarm} provide valuable abstractions for multi-agent pipeline control. 

\rebuttal{More recently, hierarchical frameworks like OWL \cite{hu2025owl} deploy domain-agnostic planners with specialized workers using task decomposition and reactive adaptive, while AgentOrchestra \cite{zhang2025agentorchestra} unifies tools, environments, and agents via TEA Protocol for agent-level coordination. Cognitive Kernel-Pro \cite{fang2025cognitive} focuses on cognitive architecture and memory for knowledge integration, and OAgents \cite{zhu2025oagents} provides a modular infrastructure for agent design and evaluation. For runtime adaptation, Co-Sight \cite{zhang2025co} embeds replanning rules in its agentic planner, smolagents \cite{smolagents} employs interval-based replanning up to fixed step limits, and dynamic workflow systems \cite{gao2025flowreasoner, nie2025weak} enable intelligent agent resource allocation. }

\rebuttal{However, these systems usually operate at agent-level granularity without research-specific optimizations. \method uniquely addresses this with the tree-structured formulation and nodel-level orchestration, enabling finer-grained resource reallocation capabilities like parallel branching, speculative execution, and early termination of redundant paths. This allows \method to dynamically adapt the research workflow in response to the evolving information with maximized efficiency.}
\begin{figure*}[t]
    \centering
    \begin{subfigure}[t]{0.59\textwidth}
        \centering
        \includegraphics[height=1.8in]{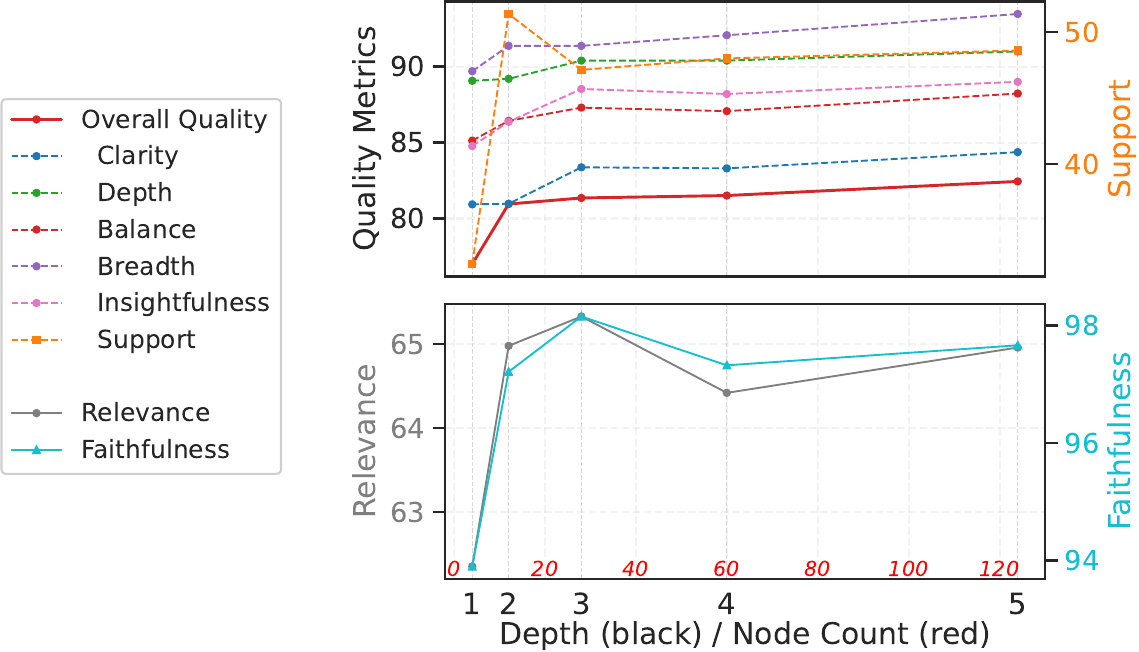}
        \caption{Depth trade-off}
        \label{fig:pilot_study_depth}
    \end{subfigure}
    \begin{subfigure}[t]{0.4\textwidth}
        \centering
        \includegraphics[height=1.8in]{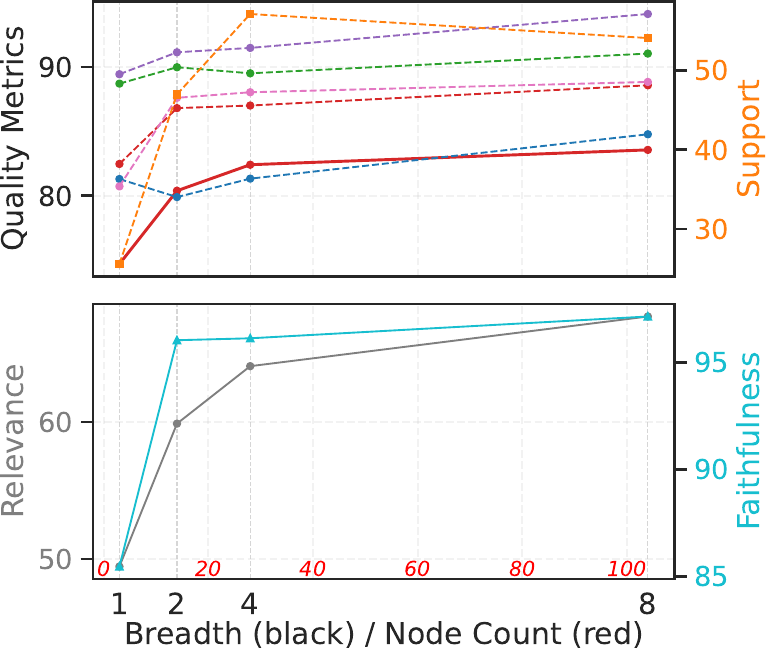}
        \caption{Breadth trade-off}
        \label{fig:pilot_study_breadth}
    \end{subfigure}
    \caption{\textbf{Trade-offs between deep research tree structure and response quality.} Left figure (a) varies \emph{depth} (breadth fixed at 4) and right figure (b) varies \emph{breadth} (depth fixed at 3); in each, the \emph{top} plot shows \emph{Quality} metrics with sub-metric \emph{Support} on the right y-axis, while the \emph{bottom} plot shows \emph{Relevance} (left) and \emph{Faithfulness} (right). The red labels along the x-axis give total node counts as a proxy for computational cost. Early increases raise quality, but gains saturate as cost escalates.}
    \label{fig:pilot_study_combined}
\end{figure*}

\section{Background}
\subsection{Problem Formulation}
Deep research tasks involve tackling complex, open-ended queries by iterative reasoning, gathering diverse information, and synthesizing knowledge into comprehensive responses. 

We formalize such a task as follows: a user query $q \in \mathcal{Q}$, where $\mathcal{Q}$ is the space of natural language queries. The goal is to produce a response $r \in \mathcal{R}$ by integrating knowledge from retrieved context $C = \{c_1, c_2, \dots, c_n\}$ sourced from a corpus $D$ (e.g., web searches or local documents). During this process, research findings $F = \{f_1, f_2, \dots, f_m\}$, comprising reasoning traces and summarized insights, are iteratively derived from the context.

Consequently, the response is generated as
\begin{equation}
r = \sigma(q, C, F),
\end{equation}
where $\sigma: \mathcal{Q} \times 2^{C} \times 2^{F} \to \mathcal{R}$ is a synthesis function that aggregates and processes the inputs to maximize quality metrics like factual accuracy, comprehensiveness, and relevance. Here, $2^{C}$ and $2^{F}$ denote the power sets of $C$ and $F$, representing all possible subsets of contexts and findings. In practice, these are subsets collected during the research process, and $\sigma$ is typically implemented with an LLM. 

To solve such tasks scalably, a deep research framework must balance thorough exploration with computational efficiency. Traditional sequential pipelines like linear retrieval-augmented generation (RAG) often struggle with complex queries, incurring high costs due to repetitive traversals or premature convergence to suboptimal research direction. Given the hierarchical and multi-faceted nature of deep research, it is natural to model the process as a tree structure, like in Tree of Thoughts \cite{yao2023tree}.

Formally, we model the process as a directed tree $\mathcal{T} = (N^P \cup N^R, E)$ with disjoint node sets of \emph{planning nodes} $N^P$ and \emph{research nodes} $N^R$. 

A planning node $n^P$ decomposes a query $q^n$ into a finite set of subqueries:
\begin{equation}
n^P(q^n) \rightarrow \{q^n_1,\ldots,q^n_{b_n}\}, \quad q^n_j \in \mathcal{Q}.
\end{equation}
Here, $q^n$ can be either the initial query or a previously decomposed subquery. The tree root is therefore the planning node $n^P_0$ that receives the initial query $q$. The number of decomposed subqueries $b_n = |n^P(q^n)|$ is defined as the \emph{breadth} at this research level. 

Each subquery $q^n_j$ further instantiates a research node $n^R(q^n_j)$, which retrieves information from the corpus $D$ and performs reasoning:
\begin{equation}
n^R(q^n_j) \rightarrow (C_{q^n_j}, F_{q^n_j}),
\end{equation}
producing a set of local context $C_{q^n_j} \subseteq C$ and research findings $F_{q^n_j} \subseteq F$. 

Optionally, a research node can be deepened by spawning a child planning node that further decomposes $q^n_j$, after which the alternating process of planning and research continues recursively. The \emph{depth} $d$ of the tree is defined as the number of research-node layers along the longest root-to-leaf path. 

The final response $r_{\mathcal{T}}$ can then be synthesized as
\begin{equation}
    r_{\mathcal{T}} = \sigma\left(q, \bigcup_{n_i \in N^R}C_i, \bigcup_{n_i \in N^R}F_i\right),
\end{equation}
by aggregating each research node's local contexts and findings across the tree. The response quality can be subsequently measured by an evaluation function $E(r)$.

However, fixing the depth and breadth of the tree can lead to suboptimal performance: shallow trees might insufficiently explore complex topics, while overly deep or wide trees can waste substantial resources on simple tasks. Therefore, the core challenge is to orchestrate the tree structure at runtime to maximize the response quality within a time budget $t_{\text{max}}$. 

Formally, we aim to solve the optimization problem:
\begin{equation}
\argmax_{\mathcal{T}} E(r_{\mathcal{T}}) \quad \text{s.t.} \quad t(\mathcal{T}) \leq t_{\text{max}},
\end{equation}
where $t(\mathcal{T})$ denotes the end-to-end latency of the research tree including all nodes and edges.

\subsection{Motivating Experiments}

Following the above formulation, we investigate how the tree structure can affect response quality in deep research tasks. We evaluate an existing deep research framework, GPT-Researcher \cite{gptresearcher}, on a set of 100 complex queries randomly sampled from DeepResearchGym \cite{coelho2025deepresearchgym}. We varied the research tree's maximum depth and breadth hyperparameters and measured performance across multiple metrics.

\paragraph{Depth} In Figure \ref{fig:pilot_study_combined}(a), increasing depth from 1 to 2 yields the largest gain, with overall quality score rising sharply from 77.00 to 80.95. Beyond depth 3, the curves flatten. Extra depth produces only marginal quality improvements while node number grows exponentially. Notably, \emph{Relevance} and \emph{Faithfulness} peak at depth 3 and then decline, as deeper searches bring in peripheral sources and redundant materials, diluting core evidence and complicating the write-up compression.

\paragraph{Breadth} A similar pattern can be observed when varying the breadth. In Figure \ref{fig:pilot_study_combined}(b), widening the tree from breadth 1 to 2 delivers a substantial quality gain with a moderate increase in nodes. Quality continues to improve up to breadth 4, after which the gains taper off. 

To summarize, initial increases in depth or breadth are valuable, but returns diminish as node counts escalate. This highlights that a one-size-fits-all approach is inefficient. Adaptive planning is essential to tailor depth and breadth to each query's complexity, optimizing the quality-cost tradeoff.

\section{\method}
\method consists of three core components: (1) an \textbf{Adaptive Research Planner}, (2) a \textbf{Runtime Orchestration Layer}, and (3) a \textbf{Fully-Asynchronous Execution Infrastructure}. Together, these components dynamically expand and prune the research tree $\mathcal{T}$ in real time and execute subtasks concurrently.

\subsection{Adaptive Research Planning}
To efficiently navigate vast information spaces, it is essential to adapt the breadth and depth of research to the scope, nature, and complexity of each query. For example, broad queries like \textit{``What is the impact of climate change?''} can be decomposed into multiple subqueries that address distinct aspects. In contrast, more specific questions like \textit{``What’s the process for developing film in a darkroom?''} require less exploration but demand greater focus and precision.

\begin{figure*}[t]
    \centering
    \includegraphics[width=.9\textwidth]{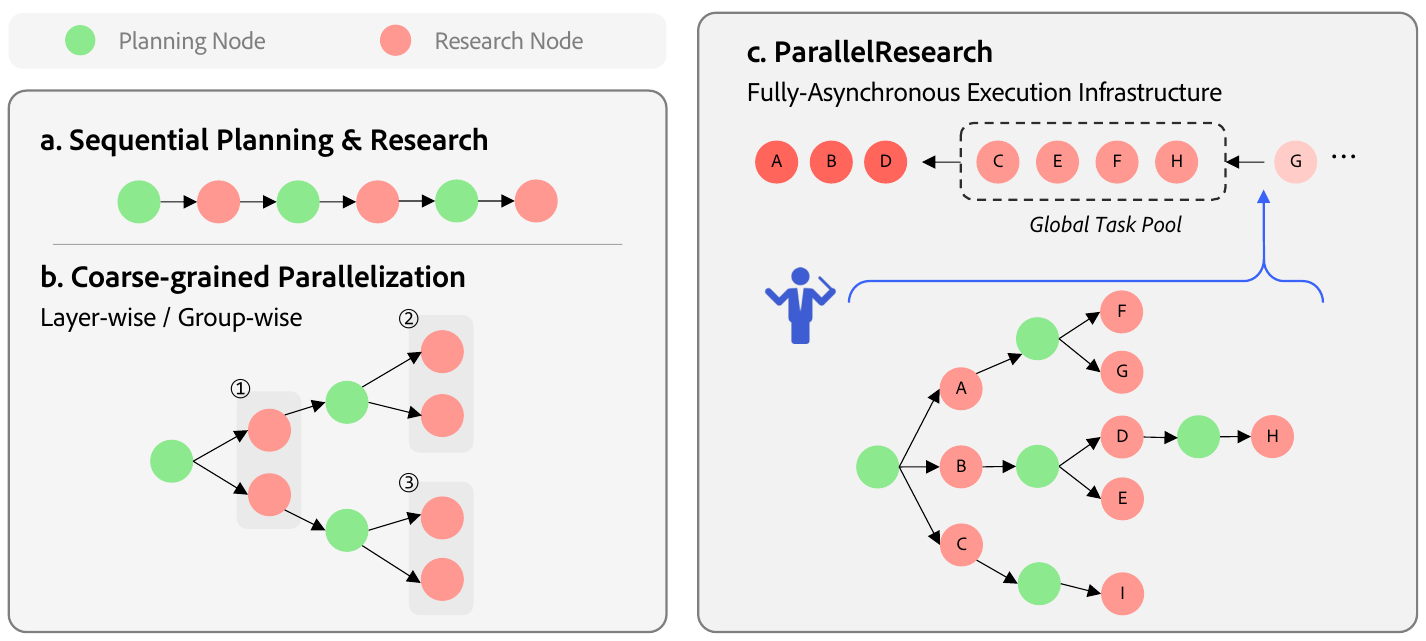}
    \caption{Sequential processing and coarse-grained parallelization at layer or group level force nodes to wait for slow dependencies and introduce unnecessary latency. \method enables fully asynchronous execution by submitting task nodes to a global pool; child nodes (e.g., \texttt{D}, \texttt{E}, \texttt{F}) can start as soon as their parents (\texttt{A}, \texttt{B}) finish, without waiting for unrelated nodes such as \texttt{C}.}
    \label{fig:parallel}
\end{figure*}

Therefore, we propose an adaptive research planner that decomposes the query $q^n$ into $b_n$ subqueries at each planning node $n^P \in N^P$, conditioned on the current query $q^n$ and the accumulated findings $F$.  Concretely, the planner applies a breadth policy $\pi_b$:
\begin{equation}
\label{eq:breadth}
n^P(q^n) = \pi_b(q^n, F) = \big(b_n, \{q^n_1, \dots, q^n_{b_n}\}\big),
\end{equation}
where $b_n$ denotes the branching factor (i.e., the number of subqueries) and $\{q^n_1, \dots, q^n_{b_n}\}$ are the generated subqueries. The key decision is to choose $b_n$ to balance the expected utility of broader exploration against the additional computation incurred by executing the spawned research nodes:
\begin{equation}
b_n
= \argmax_{b \in [1, b_{\text{max}}]}
\ \mathbb{E}\Big[\Delta U_{\text{IG}}(b \mid q^n, F)\;-\;\lambda \sum_{j=1}^{b}\Delta t\big(n^R(q^n_j)\big) \Big],
\end{equation}
where $U_{\text{IG}}(b \mid q^n, F)$ estimates the expected \emph{information gain} \cite{quinlan1986induction} from decomposing into $b$ subqueries, $t\big(n^R(q^n_j)\big)$ denotes the expected latency of running the corresponding research node, and $\lambda>0$ controls the tradeoff between response quality and computational overhead.

Once a research node finishes, a depth policy $\pi_d$ decides whether to further deepen the current branch based on the local query $q^i$ and findings $F_i$ at each research node $n_i^R \in N^R$. Specifically, $\pi_d$ compares the expected information gain from one additional research-node layer to the expected additional latency incurred, and deepens only when the gain-per-latency ratio exceeds a threshold:
\begin{equation}
\label{eq:depth}
\pi_d(q^i, F_i)
= \mathbb{I}\left\{
\frac{\mathbb{E}\!\left[\Delta U_{\text{IG}}(q^i, F_i)\right]}
{\mathbb{E}\!\left[\Delta t(q^i, F_i)\right] + \epsilon}
> \tau
\right\},
\end{equation}
where $\Delta U_{\text{IG}}(q^i, F_i)$ denotes the expected \emph{information gain} from deepening the current branch, $\Delta t(q^i, F_i)$ denotes the corresponding expected incremental latency, $\epsilon>0$ is a small constant for numerical stability, and $\tau$ is a threshold that controls diminishing returns. The output is a binary decision whether to deepen the current research path.

In our work, $\pi_b$ and $\pi_d$ are implemented with LLMs for decision-making. However, the proposed formulation is policy-agnostic and can be also instantiated using hand-crafted heuristics or learned models, \eg via reinforcement learning.

\begin{algorithm}[t]
\caption{Runtime Orchestration}
\label{alg:orchestration}
\begin{algorithmic}[1]
\Require Hyperparameters $\Phi_{\text{min}}$, $\Psi_{\text{min}}$
\Function{Orchestrator}{$n^R_i$, $q^i$, $C_i$, $F_i$}
    \State $n^R_i.\text{terminate} \gets \texttt{False}$ \textcolor{purple}{\Comment{Initialization}}
    \State \textbf{Async} Execute node $n^R_i$, update $C_i$, $F_i$ and parent
    \State \textbf{Async} Plan child queries 
    \For{each child query $q^j$}
        \State \textbf{Async} Speculatively execute $n^R_j$ with $C_j$, $F_j$    
        \State \textbf{Async} \Call{Orchestrator}{$n^R_j$, $q^j$, $C_j$, $F_j$}
    \EndFor

    \Statex \hspace{1.5em} \textcolor{purple}{$\triangleright$ Continuous monitor at each level}
    \While{not $n^R_i.\text{terminate}$}
        \State Update $C_i$, $F_i$ from $n^R_i$ and descendants $n^R_j$
        \State Pass $C_i$, $F_i$ to parent node
        \State \textbf{Async} Evaluate $(\phi_i, \psi_i) \gets \pi_o(q^i, C_i, F_i)$ 
        \If{$\phi_i \geq \Phi_{\text{min}}$ and $\psi_i \geq \Psi_{\text{min}}$}
            \State $n^R_i.\text{terminate} \gets \texttt{True}$
            \State Kill node $n^R_i$ \textcolor{purple}{\Comment{Early termination}}
            \For{each descendant $n^R_j$}  \textcolor{purple}{\Comment{Pruning}}
                \State Kill $n^R_j$ and descendants recursively
            \EndFor
        \EndIf
        \If{$n^R_i$ and all children finished/killed}
            \State $n^R_i.\text{terminate} \gets \texttt{True}$
        \EndIf
    \EndWhile
\EndFunction
\end{algorithmic}
\end{algorithm}

\subsection{Runtime Orchestration}
To address the dynamic and evolving nature of research, where priorities can shift as new evidence emerges, breadth planning conducted \emph{prior-research} and depth planning performed \emph{post-research} can be limiting. Furthermore, depth planning can also delay the exploitation of promising research paths until decisions are finalized. 

To mitigate these, we introduce a runtime orchestration layer that dynamically manages the research tree $\mathcal{T}$ using \emph{mid-research} signals for resource reallocation. Each research node $n^R(q^i)$ is continuously monitored by an orchestration policy $\pi_o$ conditioned on the local query $q^i$, real-time context $C_i$, and accumulated findings $F_i$:
\begin{equation}
\label{eq:monitor}
\pi_o(q^i, C_i, F_i) = (\phi_i, \psi_i),
\end{equation}
where $\phi_i \in [0,1]$ denotes a goal-satisfaction score and $\psi_i \in [0,1]$ denotes a quality score. The orchestration layer uses these signals to determine whether to continue scheduling work on node $i$:
\begin{equation}
\label{eq:schedule_rule}
\textsc{Continue}(i) = \mathbb{I}\{\phi_i \ge \Phi_{\min} \ \wedge\  \psi_i \ge \Psi_{\min}\},
\end{equation}
where $\Phi_{\min}$ and $\Psi_{\min}$ are thresholds for goal satisfaction and quality, respectively. In our implementation, $\pi_o$ is instantiated with an LLM.

More importantly, this mechanism enables \emph{speculative execution}: child nodes can be spawned and deepen the tree without awaiting the parent's planning decision. Child nodes' findings update the parent's $C_i$ and $F_i$, even after the parent's research completes, enabling recursive and adaptive task management. As shown in Algorithm \ref{alg:orchestration}, upon evaluation at each hierarchy, low-yield nodes and their descendants are terminated early once the research goal is satisfied, pruning the subtree dynamically.

\subsection{Fully-Asynchronous Execution Infrastructure}
To maximize efficiency in traversing the evolving research tree $\mathcal{T}$, \method incorporates a fully-asynchronous execution infrastructure that enables concurrent processing across multiple axes: breadth (parallel subqueries at the same level), depth (speculative deepening of paths), and across the runtime orchestrators at different recursive hierarchies in Algorithm \ref{alg:orchestration}. 

The engine operates by submitting all research nodes $n^R_i$ to a global asynchronous task pool as soon as they are planned and orchestrated. Each node is parameterized by its local query $q^i$, depth $d_i$, parent identifier, and a unique status identifier. Dependencies are enforced dynamically: a child node $n^R_j$ becomes eligible for execution only once its parent $n^R_i$ completes its initial research phase, but speculative spawning allows planning and partial execution to begin earlier under the runtime orchestrator's guidance.

\method's engine leverages non-blocking asynchronous calls, allowing tasks to progress independently. This approach mitigates bottlenecks inherent in sequential or coarse-grained parallel execution, where nodes must wait for unrelated dependencies to complete. For instance, as illustrated in Figure \ref{fig:parallel}, child nodes (e.g., \texttt{D}, \texttt{E}, \texttt{F}) can initiate immediately upon their respective parents' (\texttt{A}, \texttt{B}) completion, without delays from slower siblings like \texttt{C}. 
\section{Experiments}

\begin{table*}[t]
  \centering
  \tiny
  \setlength{\tabcolsep}{4pt}
  \caption{Overall evaluation results on DeepResearch Bench under flexible time budgets, judged by \textbf{\texttt{Gemini-2.5-flash}} and \textbf{\texttt{Gemini-2.5-pro}}. \method(-AP, -RO) refers to the ablation without adaptive planning and runtime orchestration. Commercial deep research agents' results are directly from the DeepResearch Bench Leaderboard\protect\footnotemark with no latency statistics reported. }  
  \label{tab:bench}
\renewcommand{\arraystretch}{1.1}
 \resizebox{\textwidth}{!}{
  \begin{tabular}{l *{9}{c}}
  \toprule
  \multirow{2}{*}{\textbf{Method}}
    & \multicolumn{2}{c}{\textbf{Throughput}} & \multicolumn{5}{c}{\textbf{RACE}}
    & \multicolumn{2}{c}{\textbf{FACT}} \\ \cmidrule(lr){2-3}
  \cmidrule(lr){4-8}\cmidrule(lr){9-10}
    & \textbf{\# Nodes} & \textbf{Latency} &\textbf{Overall} & \textbf{Comp.} & \textbf{Depth}
    & \textbf{Inst.} & \textbf{Read.}
    & \textbf{Cit. Acc.} & \textbf{Eff. Cit.} \\
  \midrule
    Grok Deeper Search & - & - & 38.22 & 36.08 & 30.89 & 46.59 & 42.17 & 73.08 & 8.58 \\
    Perplexity Research & - & - & 40.46 & 39.10 & 35.65 & 46.11 & 43.08 & 82.63 & 31.20 \\
    OpenAI Deep Research & - & - & 46.45 & 46.46 & 43.73 & 49.39 & 47.22 & 75.01 & 39.79\\
    Gemini-2.5-Pro Deep Research & - & - & 49.71 & 49.51 & 49.45 & 50.12 & 50.00 & 78.30 & 165.34\\
    \midrule
    GPT-Researcher & 23.12 & 554.41 s & 41.15
    & 38.58 & 37.55 & 46.03 & 45.62 & 65.58 & 9.40 \\
    \ablation & 27.88 & \textbf{207.06} s & 41.33 
    & 38.61 & 38.09 & 46.01 & 45.80 & 70.06 & 17.35 \\
   \method & \textbf{39.30} & 367.88 s & \textbf{41.92} 
   & \textbf{39.55} & \textbf{38.61} & \textbf{46.36} & \textbf{45.83}
    & 58.25 & 22.94 \\
  \bottomrule
  \end{tabular}}
  \vspace{-1em}
\end{table*}

\begin{table*}[t!]
  \centering
  \setlength{\tabcolsep}{4.0pt}
  \caption{Evaluation of deep research frameworks on DeepResearchGym under fixed time budgets. Scores are assessed by \textbf{\texttt{gpt-4.1-mini-2025-04-14}} \rebuttal{and averaged over 5 runs. All metrics reported with 95\% confidence intervals.}}
  \label{tab:gym}
  \renewcommand{\arraystretch}{1.15}
  \resizebox{\textwidth}{!}{\begin{tabular}{l c ccccccc cc c}
    \toprule
    & \textbf{Throughput} & \multicolumn{7}{c}{\textbf{Quality}} 
    & \multicolumn{2}{c}{\textbf{Relevance}} 
    & \textbf{Faithfulness} \\ \cmidrule(lr){2-2}
    \cmidrule(lr){3-9} \cmidrule(lr){10-11} \cmidrule(lr){12-12}
     & \# Nodes & \textbf{Overall} & Clarity & Depth & Balance & Breadth & Support & Insight
    & KPR & KPC ($\downarrow$) 
     & Cit. Recall \\
\midrule
\textbf{\underline{2 minutes}}\\
    GPT-Researcher
    & 8.00 ± 0.18
    & 74.06 ± 0.88 & 78.66 ± 1.94 & 88.88 ± 0.70 & 84.70 ± 0.80 & 90.64 ± 0.27 & 21.22 ± 3.39 & 80.26 ± 1.88
    & 54.50 ± 2.16 & 0.76 ± 0.35
    & 85.54 ± 1.77 \\
    \method
    & 19.42 ± 1.31
    & \textbf{79.78 ± 0.80} & \textbf{83.88 ± 1.06} & 90.26 ± 0.22
    & \textbf{87.36 ± 0.47} & 91.10 ± 0.27 & \textbf{40.10 ± 3.91} & 85.98 ± 1.01
    & 62.29 ± 2.00 & 0.73 ± 0.26
    & \textbf{94.71 ± 0.87} \\
    \method (-AP)
    & 8.94 ± 0.97
    & 79.44 ± 0.76 & 86.64 ± 0.68 & \textbf{90.58 ± 0.24} & 87.26 ± 0.72 & \textbf{92.22 ± 0.39} & 33.80 ± 3.88 & 86.16 ± 1.15
    & 62.73 ± 1.99 & 0.77 ± 0.25
    & 84.88 ± 1.78 \\
    \method (-AP, -RO)
    & \textbf{21.14 ± 1.01} 
    & 78.88 ± 0.96 & 82.10 ± 1.37 & 89.04 ± 0.75 & 85.36 ± 1.02 & 89.94 ± 0.92 & 39.62 ± 3.91 & \textbf{87.22 ± 0.86} 
    & \textbf{66.72 ± 2.01} & \textbf{0.43 ± 0.21}
    & 94.36 ± 0.86 \\
\midrule
\textbf{\underline{10 minutes}}\\   
    GPT-Researcher
    & 23.94 ± 0.83
    & 79.52 ± 0.80 & 82.78 ± 1.27 & 89.96 ± 0.41 & 86.60 ± 0.51 & 90.64 ± 0.44 & 39.48 ± 3.91 & 87.64 ± 0.78
    & 63.88 ± 1.88 & 0.61 ± 0.27
    & 95.53 ± 0.80 \\
    \method
    & \textbf{98.43 ± 0.19} 
    & \textbf{83.90 ± 0.78} & 83.08 ± 1.25 & 90.52 ± 0.43 & \textbf{88.08 ± 0.55} & \textbf{94.42 ± 0.44} & \textbf{58.10 ± 3.81} & 89.18 ± 0.46 
    & 66.16 ± 1.93 & 0.55 ± 0.22
    & 95.96 ± 1.03 \\
    \method (-AP)
    & 62.32 ± 9.84 
    & 83.12 ± 0.79 & 82.80 ± 1.30 & \textbf{90.60 ± 0.31} & 87.86 ± 0.59 & 93.88 ± 0.43 & 54.20 ± 3.94 & \textbf{89.38 ± 0.25}
    & \textbf{68.42 ± 1.91} & \textbf{0.56 ± 0.23}
    & \textbf{97.57 ± 0.49} \\
    \method (-AP, -RO)
    & 68.00 ± 0.00 
    & 82.69 ± 1.04 & \textbf{83.28 ± 1.32} & 89.42 ± 0.95 & 87.44 ± 0.87 & 92.46 ± 0.76 & 56.24 ± 3.85 & 87.32 ± 1.14 
    & 66.44 ± 1.98 & 0.51 ± 0.21
    & 96.52 ± 0.65 \\
    \bottomrule
  \end{tabular}}
\end{table*}

\subsection{Benchmarks and Evaluation Metrics}
We evaluate our method on two deep research benchmarks.

\textbf{DeepResearch Bench} \cite{du2025deepresearch} comprises 100 PhD-level research tasks across 22 distinct fields. These tasks were designed by domain experts based on a statistical analysis of over 96,000 real-world user queries from search-enabled LLM interactions. We use the whole English subset with 50 questions for evaluation. The benchmark proposes two evaluation frameworks: RACE for report quality and FACT for citation trustworthiness.
\begin{itemize}
\itemsep0em
\vspace{-.5em}
\item \textbf{RACE}: 
\begin{itemize}
\vspace{-.2em}
\itemsep0em
\item \textbf{Comprehensiveness}: Evaluates thorough coverage of relevant aspects, including diverse perspectives and key subtopics, ensuring understanding without omissions.
\item \textbf{Depth}: Assesses level of detail, analysis, and insights beyond surface-level, including causes, impacts, and trends.
\item \textbf{Instruction following}: Checks adherence to query requirements, ensuring alignment with intent by following the topic and answering directly.
\item \textbf{Readability}: Assesses clarity through structure, language, and ease of understanding.
\end{itemize}
\item \textbf{FACT}:
\begin{itemize}
\vspace{-.2em}
\itemsep0em
\item \textbf{Effective citation count}: Measures factual abundance by counting the number of unique, relevant citations that effectively support key statements in the report. 
\item \textbf{Citation accuracy}: Evaluates citation trustworthiness by assessing the proportion of citations that effectively support the referenced claims. 
\end{itemize}
\end{itemize}
\footnotetext{\url{https://huggingface.co/spaces/Ayanami0730/DeepResearch-Leaderboard}}
\textbf{DeepResearchGym} \cite{coelho2025deepresearchgym} is an open-source evaluation sandbox for deep research systems. The benchmark consists of 1,000 complex, high-engagement queries from the Researchy Questions dataset \cite{rosset2025researchy}. Due to resource constraints, we randomly sampled 100 for evaluation. \rebuttal{To ensure representativeness, we performed rigorous statistical validation with Mann-Whitney U, Kolmogorov-Smirnov, and equivalence testing in Appendix~\ref{app:sample}, Table~\ref{tab:statistical_tests}, showing no significant distributional differences between our sample and the full dataset across all feature dimensions ($p>0.05$, Cohen's d ranging from $-0.234$ to $0.070$).} It employs an LLM-as-a-judge protocol to assess generated reports along three dimensions:
\begin{itemize}
\vspace{-0.5em}
    \item \textbf{Quality:} Measures the report quality with a rubric of fine-grained criteria, including clarity, depth, balance, breadth, insightfulness, and support. 
     \item \textbf{Relevance:} Measures how well the generated reports satisfy the information needs.
     \item \textbf{Faithfulness:} Measures the report and citation factuality.
\end{itemize}

\subsection{Evaluation Setup}
To ensure a fair and controlled evaluation where performance differences can be attributed solely to our runtime orchestration, we implement \method on top of GPT-Researcher system \cite{gptresearcher} and treat the original GPT-Researcher as baseline. Although our approach is system-agnostic, adopting a single, stable baseline allows us to isolate the effect of orchestration while keeping all other components fixed. We use the search API from DeepResearchGym across all experiments for consistency.

On DeepResearch Bench, we allow the system to freely execute with a maximum breadth and depth. On DeepResearchGym, we set maximum execution times at 2 and 10 minutes to reflect realistic usage scenarios:
\begin{itemize}
    \vspace{-.5em}
    \item The 2-minute cutoff reflects human multitasking behavior, where information workers spend an average of $\sim$2--3 minutes on events or tools before task switching \cite{gonzalez2004constant}.
    \item The 10-minute threshold matches the average duration of a ``working sphere'' \cite
    {gonzalez2004constant} and is supported by evidence from high-performance computing \cite
    {schlagkamp2015acceptance} and crowdsourcing tasks \cite{bernstein2011crowds}, indicating that a 10-minute  window preserves task continuity without slowdown.
\end{itemize}

\subsection{Results}

\paragraph{DeepResearch Bench.} To assess our system's performance relative to other deep research agents, we evaluate it under flexible time budgets on DeepResearch Bench. As shown in Table~\ref{tab:bench}, \method achieves substantial efficiency gains, processing 39.3 nodes on average while reducing latency by 1.51$\times$ compared to the baseline. 

In terms of quality, our proposed framework consistently improves across all RACE sub-metrics. Compared to the ablation without adaptive planning and runtime orchestration, \method incurs higher latency but conducts more research at a comparable throughput, highlighting a favorable trade-off between efficiency and comprehensiveness. Importantly, \method also achieves a performance competitive with commercial systems like Grok Deeper Search and Perplexity Research, narrowing the gap with proprietary agents. Beyond quantitative metrics, we also provide a detailed case analysis of \method's adaptability across diverse scenarios in Appendix \ref{app:case}.

\paragraph{DeepResearchGym.} Under fixed time budgets, \method consistently delivers superior throughput, processing substantially more research than the GPT-Researcher baseline (up to 4.11$\times$ in the 10-minute setup), as shown in Table \ref{tab:gym}. \rebuttal{Per-component ablations with 95\% confidence intervals isolate the contribution of each mechanism, demonstrating statistically significant gains from adaptive planning and runtime orchestration, which enable speculative execution without compromising quality.}

Beyond throughput, \method also improves overall response quality, with clear improvements in balance, breadth, and insight metrics. These gains highlight its ability to maintain a robust trade-off between expansive coverage and focused analysis, particularly under tight time constraints. Notably, the overall quality of \method with 2-minute execution even surpasses that of the GPT-Researcher baseline with 10 minutes, demonstrating a 5$\times$ speed-up while preserving quality.

\rebuttal{To validate generalizability, we also experiment on open-source \texttt{Qwen3-235B-A22B-Instruct}. Results in Appendix Table~\ref{tab:new_results} indicate that \method achieves a consistent 5$\times$ quality-preserving speedup, confirming that our efficiency improvements transfer across model families.} 

Overall, these results underscore \method's capacity for efficient deep research across diverse benchmarks, model families, LLM judges, and time constraints.

\section{Analysis}
\rebuttal{To provide deeper insights into \method's robustness and practical viability, we conducted comprehensive analyses on evaluation reliability, failure patterns, and economic efficiency.}

\paragraph{Evaluation Reliability.}
\rebuttal{To validate the reliability of LLM-as-a-judge evaluations, we conducted inter-rater analysis across five independent runs in Appendix~\ref{app:llm_reliability}, Table~\ref{tab:icc_scores}. The Intraclass Correlation Coefficient (ICC) scores are consistently above 0.85 (mostly $>$0.9) across all metrics and systems, indicating excellent consistency and reliability of the LLM judges. This confirms that our evaluation protocol can produce stable and trustworthy assessments.}

\paragraph{Failure Analysis.}
\rebuttal{To understand when \method excels or struggles, we analyzed performance patterns across different query types in Appendix~\ref{app:patterns}, Figure~\ref{fig:patterns}. \method achieves a 70\% win rate overall, with superior performance on knowledge-intensive and reasoning-intensive queries where adaptive planning effectively balances breadth and depth. These query types benefit most from dynamic resource allocation across multiple research branches. However, performance limitations emerge primarily for ambiguous or incomplete queries where the optimal decomposition strategy is unclear, suggesting opportunities for future improvements in intent clarification mechanisms.}

\paragraph{Economic Cost Analysis.}
\rebuttal{Cost profiling in Appendix~\ref{app:cost}, Table~\ref{tab:token_usage} shows orchestration overhead accounts for only 7\% of total LLM API costs, while early pruning saves 3.39$\times$ more than orchestration costs. A 2-minute \method run costs \$0.21 and outperforms a 10-minute baseline costing \$0.39. Sensitivity analysis (Appendix~\ref{app:sensitivity}, Table~\ref{tab:time_budget_significance}) confirms \method maintains efficiency at 2 minutes with strategic time utilization, achieving strong gains in Support metrics (+16--18, p$<$0.001) versus the baseline's less efficient sensitivity.}

\section{Conclusion}
By formulating efficient deep research as a tree-structured optimization problem and exploiting parallelism across nodes, \method achieves significant improvements in research throughput and response quality through adaptive planning, runtime orchestration, and a fully-asynchronous execution infrastructure. Future work includes incorporating richer modalities beyond text and exploring tighter integration with human-in-the-loop feedback to further improve policies' decision-making.

\section*{Ethics Statement}
This work does not involve private data or sensitive information. Experiments were conducted using publicly available resources. While our framework aims to improve the efficiency and quality of deep research systems, we acknowledge the broader risks of misuse, including the potential amplification of biased or unreliable information. Responsible deployment requires careful selection of data sources, robust fact-checking, and adherence to ethical standards.

\bibliography{iclr2026_conference}

@article{coelho2025deepresearchgym,
  title={Deepresearchgym: A free, transparent, and reproducible evaluation sandbox for deep research},
  author={Coelho, Jo{\~a}o and Ning, Jingjie and He, Jingyuan and Mao, Kangrui and Paladugu, Abhijay and Setlur, Pranav and Jin, Jiahe and Callan, Jamie and Magalh{\~a}es, Jo{\~a}o and Martins, Bruno and others},
  journal={arXiv preprint arXiv:2505.19253},
  year={2025}
}

@article{zhang2025evoflow,
  title={EvoFlow: Evolving Diverse Agentic Workflows On The Fly},
  author={Zhang, Guibin and Chen, Kaijie and Wan, Guancheng and Chang, Heng and Cheng, Hong and Wang, Kun and Hu, Shuyue and Bai, Lei},
  journal={arXiv preprint arXiv:2502.07373},
  year={2025}
}

@article{zhang2024aflow,
  title={Aflow: Automating agentic workflow generation},
  author={Zhang, Jiayi and Xiang, Jinyu and Yu, Zhaoyang and Teng, Fengwei and Chen, Xionghui and Chen, Jiaqi and Zhuge, Mingchen and Cheng, Xin and Hong, Sirui and Wang, Jinlin and others},
  journal={arXiv preprint arXiv:2410.10762},
  year={2024}
}

@article{niu2025flow,
  title={Flow: A Modular Approach to Automated Agentic Workflow Generation},
  author={Niu, Boye and Song, Yiliao and Lian, Kai and Shen, Yifan and Yao, Yu and Zhang, Kun and Liu, Tongliang},
  journal={arXiv preprint arXiv:2501.07834},
  year={2025}
}

@Misc{gptresearcher,
author = {Elovic, Assaf},
month = jul,
title = {{GPT Researcher: LLM based autonomous agent that conducts deep local and web research on any topic and generates a long report with citations}},
url = {https://github.com/assafelovic/gpt-researcher},
version = {0.5.4},
year = {2023}
}

@article{alzubi2025open,
  title={Open deep search: Democratizing search with open-source reasoning agents},
  author={Alzubi, Salaheddin and Brooks, Creston and Chiniya, Purva and Contente, Edoardo and von Gerlach, Chiara and Irwin, Lucas and Jiang, Yihan and Kaz, Arda and Nguyen, Windsor and Oh, Sewoong and others},
  journal={arXiv preprint arXiv:2503.20201},
  year={2025}
}

@Misc{langchain,
  title =        {Open Deep Research},
  author =       {Langchain},
  url = {https://github.com/langchain-ai/open_deep_research},
  year =         {2025}
}

@article{pan2025specreason,
  title={Specreason: Fast and accurate inference-time compute via speculative reasoning},
  author={Pan, Rui and Dai, Yinwei and Zhang, Zhihao and Oliaro, Gabriele and Jia, Zhihao and Netravali, Ravi},
  journal={arXiv preprint arXiv:2504.07891},
  year={2025}
}

@article{autogen,
  title={Autogen: Enabling next-gen llm applications via multi-agent conversation framework},
  author={Wu, Qingyun and Bansal, Gagan and Zhang, Jieyu and Wu, Yiran and Zhang, Shaokun and Zhu, Erkang and Li, Beibin and Jiang, Li and Zhang, Xiaoyun and Wang, Chi},
  journal={arXiv preprint arXiv:2308.08155},
  volume={3},
  number={4},
  year={2023}
}

@misc{langgraph,
  title        = {LangGraph: Build resilient language agents as graphs},
  author       = {LangChain},
  url = {https://langchain-ai.github.io/langgraph/},
  year         = {2024}
}

@misc{openai_swarm,
  title        = {Swarm: Educational framework exploring ergonomic, lightweight multi-agent orchestration},
  author       = {OpenAI},
  howpublished = {\url{https://github.com/openai/swarm}},
  year         = {2024}
}

@article{dspy,
  title={DSPy: Compiling Declarative Language Model Calls into Self-Improving Pipelines},
  author={Khattab, Omar and Singhvi, Arnav and Maheshwari, Paridhi and Zhang, Zhiyuan and Santhanam, Keshav and Vardhamanan, Sri and Haq, Saiful and Sharma, Ashutosh and Joshi, Thomas T. and Moazam, Hanna and Miller, Heather and Zaharia, Matei and Potts, Christopher},
  journal={The Twelfth International Conference on Learning Representations},
  year={2024}
}

@article{miao2023specinfer,
  title={SpecInfer: Accelerating Large Language Model Serving with Speculative Inference},
  author={Miao, Xiaohui and Zuo, Senzhang and Li, Ang and Guo, Jiaqi and Zhang, Xiaoying and Xing, Yifan and Qian, Xiaoliang and Gao, Yanzhi and Zhou, Jingren and Zhou, Fan},
  journal={arXiv preprint arXiv:2305.09781},
  year={2023}
}

@article{cai2024medusa,
  title={Medusa: Simple LLM Inference Acceleration Framework with Multiple Decoding Heads},
  author={Cai, Tianle and Li, Jiayi and Geng, Haotian and Ge, Yuxin and Zhang, Shizhe and others},
  journal={arXiv preprint arXiv:2401.10774},
  year={2024}
}

@article{du2025deepresearch,
  title={DeepResearch Bench: A Comprehensive Benchmark for Deep Research Agents},
  author={Du, Mingxuan and Xu, Benfeng and Zhu, Chiwei and Wang, Xiaorui and Mao, Zhendong},
  journal={arXiv preprint arXiv:2506.11763},
  year={2025}
}

@article{nakano2021webgpt,
  title={Webgpt: Browser-assisted question-answering with human feedback},
  author={Nakano, Reiichiro and Hilton, Jacob and Balaji, Suchir and Wu, Jeff and Ouyang, Long and Kim, Christina and Hesse, Christopher and Jain, Shantanu and Kosaraju, Vineet and Saunders, William and others},
  journal={arXiv preprint arXiv:2112.09332},
  year={2021}
}

@article{yao2023tree,
  title={Tree of thoughts: Deliberate problem solving with large language models},
  author={Yao, Shunyu and Yu, Dian and Zhao, Jeffrey and Shafran, Izhak and Griffiths, Tom and Cao, Yuan and Narasimhan, Karthik},
  journal={Advances in neural information processing systems},
  volume={36},
  pages={11809--11822},
  year={2023}
}

@inproceedings{yao2023react,
  title={React: Synergizing reasoning and acting in language models},
  author={Yao, Shunyu and Zhao, Jeffrey and Yu, Dian and Du, Nan and Shafran, Izhak and Narasimhan, Karthik and Cao, Yuan},
  booktitle={International Conference on Learning Representations (ICLR)},
  year={2023}
}

@article{xu2025comprehensive,
  title={A Comprehensive Survey of Deep Research: Systems, Methodologies, and Applications},
  author={Xu, Renjun and Peng, Jingwen},
  journal={arXiv preprint arXiv:2506.12594},
  year={2025}
}

@article{yang2025speculative,
  title={Speculative thinking: Enhancing small-model reasoning with large model guidance at inference time},
  author={Yang, Wang and Yue, Xiang and Chaudhary, Vipin and Han, Xiaotian},
  journal={arXiv preprint arXiv:2504.12329},
  year={2025}
}

@article{zheng2025parallel,
  title={Parallel-R1: Towards Parallel Thinking via Reinforcement Learning},
  author={Zheng, Tong and Zhang, Hongming and Yu, Wenhao and Wang, Xiaoyang and Yang, Xinyu and Dai, Runpeng and Liu, Rui and Bao, Huiwen and Huang, Chengsong and Huang, Heng and others},
  journal={arXiv preprint arXiv:2509.07980},
  year={2025}
}

@article{ding2025dynamic,
  title={Dynamic parallel tree search for efficient llm reasoning},
  author={Ding, Yifu and Jiang, Wentao and Liu, Shunyu and Jing, Yongcheng and Guo, Jinyang and Wang, Yingjie and Zhang, Jing and Wang, Zengmao and Liu, Ziwei and Du, Bo and others},
  journal={arXiv preprint arXiv:2502.16235},
  year={2025}
}

@article{wen2025parathinker,
  title={ParaThinker: Native Parallel Thinking as a New Paradigm to Scale LLM Test-time Compute},
  author={Wen, Hao and Su, Yifan and Zhang, Feifei and Liu, Yunxin and Liu, Yunhao and Zhang, Ya-Qin and Li, Yuanchun},
  journal={arXiv preprint arXiv:2509.04475},
  year={2025}
}

@inproceedings{leviathan2023speculative,
  title={Fast Inference from Transformers via Speculative Decoding},
  author={Leviathan, Yaniv and Kalman, Matan and Matias, Yossi},
  booktitle={Proceedings of the 40th International Conference on Machine Learning},
  series={Proceedings of Machine Learning Research},
  volume={202},
  pages={19274--19286},
  year={2023},
  publisher={PMLR},
  url={https://proceedings.mlr.press/v202/leviathan23a.html}
}

@inproceedings{rosset2025researchy,
  title={Researchy Questions: A Dataset of Multi-Perspective, Decompositional Questions for Deep Research},
  author={Rosset, Corbin and Chung, Ho-Lam and Qin, Guanghui and Chau, Ethan and Feng, Zhuo and Awadallah, Ahmed and Neville, Jennifer and Rao, Nikhil},
  booktitle={Proceedings of the 48th International ACM SIGIR Conference on Research and Development in Information Retrieval},
  pages={3712--3722},
  year={2025}
}

@inproceedings{iqbal2007disruption,
  title={Disruption and recovery of computing tasks: field study, analysis, and directions},
  author={Iqbal, Shamsi T and Horvitz, Eric},
  booktitle={Proceedings of the SIGCHI conference on Human factors in computing systems},
  pages={677--686},
  year={2007}
}

@inproceedings{mark2008cost,
  title={The cost of interrupted work: more speed and stress},
  author={Mark, Gloria and Gudith, Daniela and Klocke, Ulrich},
  booktitle={Proceedings of the SIGCHI conference on Human Factors in Computing Systems},
  pages={107--110},
  year={2008}
}

@inproceedings{bernstein2011crowds,
  title={Crowds in two seconds: Enabling realtime crowd-powered interfaces},
  author={Bernstein, Michael S and Brandt, Joel and Miller, Robert C and Karger, David R},
  booktitle={Proceedings of the 24th annual ACM symposium on User interface software and technology},
  pages={33--42},
  year={2011}
}

@inproceedings{gonzalez2004constant,
  title={" Constant, constant, multi-tasking craziness" managing multiple working spheres},
  author={Gonz{\'a}lez, Victor M and Mark, Gloria},
  booktitle={Proceedings of the SIGCHI conference on Human factors in computing systems},
  pages={113--120},
  year={2004}
}

@inproceedings{schlagkamp2015acceptance,
  title={Acceptance of waiting times in high performance computing},
  author={Schlagkamp, Stephan and Renker, Johanna},
  booktitle={International Conference on Human-Computer Interaction},
  pages={709--714},
  year={2015},
  organization={Springer}
}

@article{haman2025fake,
  title={Fake no more: the redemption of ChatGPT in literature reviews},
  author={Haman, Michael and {\v{S}}koln{\'\i}k, Milan},
  journal={Accountability in Research},
  pages={1--3},
  year={2025},
  publisher={Taylor \& Francis}
}

@inproceedings{gambrell2025ai,
  title={Ai for Egovernance: Combining Artificial Intelligence and Collective Intelligence to Develop Evidence-Based Ai Policy},
  author={Gambrell, Dane},
  booktitle={2025 Eleventh International Conference on eDemocracy \& eGovernment (ICEDEG)},
  pages={317--324},
  year={2025},
  organization={IEEE}
}

@article{hu2025owl,
  title={Owl: Optimized workforce learning for general multi-agent assistance in real-world task automation},
  author={Hu, Mengkang and Zhou, Yuhang and Fan, Wendong and Nie, Yuzhou and Xia, Bowei and Sun, Tao and Ye, Ziyu and Jin, Zhaoxuan and Li, Yingru and Chen, Qiguang and others},
  journal={arXiv preprint arXiv:2505.23885},
  year={2025}
}

@article{zhang2025agentorchestra,
  title={AgentOrchestra: Orchestrating Hierarchical Multi-Agent Intelligence with the Tool-Environment-Agent (TEA) Protocol},
  author={Zhang, Wentao and Zeng, Liang and Xiao, Yuzhen and Li, Yongcong and Cui, Ce and Zhao, Yilei and Hu, Rui and Liu, Yang and Zhou, Yahui and An, Bo},
  journal={arXiv preprint arXiv:2506.12508},
  year={2025}
}

@article{fang2025cognitive,
  title={Cognitive kernel-pro: A framework for deep research agents and agent foundation models training},
  author={Fang, Tianqing and Zhang, Zhisong and Wang, Xiaoyang and Wang, Rui and Qin, Can and Wan, Yuxuan and Ma, Jun-Yu and Zhang, Ce and Chen, Jiaqi and Li, Xiyun and others},
  journal={arXiv preprint arXiv:2508.00414},
  year={2025}
}

@article{zhu2025oagents,
  title={Oagents: An empirical study of building effective agents},
  author={Zhu, He and Qin, Tianrui and Zhu, King and Huang, Heyuan and Guan, Yeyi and Xia, Jinxiang and Yao, Yi and Li, Hanhao and Wang, Ningning and Liu, Pai and others},
  journal={arXiv preprint arXiv:2506.15741},
  year={2025}
}

@article{gao2025flowreasoner,
  title={Flowreasoner: Reinforcing query-level meta-agents},
  author={Gao, Hongcheng and Liu, Yue and He, Yufei and Dou, Longxu and Du, Chao and Deng, Zhijie and Hooi, Bryan and Lin, Min and Pang, Tianyu},
  journal={arXiv preprint arXiv:2504.15257},
  year={2025}
}

@article{nie2025weak,
  title={Weak-for-Strong: Training Weak Meta-Agent to Harness Strong Executors},
  author={Nie, Fan and Feng, Lan and Ye, Haotian and Liang, Weixin and Lu, Pan and Yao, Huaxiu and Alahi, Alexandre and Zou, James},
  journal={arXiv preprint arXiv:2504.04785},
  year={2025}
}

@article{zhang2025co,
  title={Co-sight: Enhancing llm-based agents via conflict-aware meta-verification and trustworthy reasoning with structured facts},
  author={Zhang, Hongwei and Lu, Ji and Jiang, Shiqing and Zhu, Chenxiang and Xie, Li and Zhong, Chen and Chen, Haoran and Zhu, Yurui and Du, Yongsheng and Gao, Yanqin and others},
  journal={arXiv preprint arXiv:2510.21557},
  year={2025}
}

@Misc{smolagents,
  title =        {`smolagents`: a smol library to build great agentic systems.},
  author =       {Aymeric Roucher and Albert Villanova del Moral and Thomas Wolf and Leandro von Werra and Erik Kaunismäki},
  howpublished = {\url{https://github.com/huggingface/smolagents}},
  year =         {2025}
}

@article{quinlan1986induction,
  title={Induction of decision trees},
  author={Quinlan, J. Ross},
  journal={Machine learning},
  volume={1},
  number={1},
  pages={81--106},
  year={1986},
  publisher={Springer}
}
\bibliographystyle{iclr2026_conference}

\appendix
\clearpage
\onecolumn
\raggedbottom
\section{Implementation Details}
\label{app:details}

\subsection{Experimental Setups}
For model configuration, we use \texttt{gpt-4.1-mini-2025-04-14} for the main research processing, and \texttt{o3-mini-2025-01-31} for implementing the policies in Equations \ref{eq:breadth}, \ref{eq:depth} and \ref{eq:monitor} for adaptive research planning and runtime orchestration.

To ensure fair comparisons among deep research systems, we impose a maximum execution time for research trees. Once the time cut-off is reached, the research process terminates immediately, and the system generates a response based on the findings and context gathered up to that point. To allow all systems to fully utilize their time budgets, we set the maximum tree depth to 10 and the maximum breadth to 4 within the GPT-Researcher framework. To provide additional flexibility, the adaptive research planning module may expand the breadth up to 6 when necessary.

Additionally, to control the computational costs introduced by the runtime orchestration layer, we set an interval of 8 seconds between successive evaluations of goal satisfaction and research quality.

\section{Evaluation Robustness and Reliability}
\label{app:evaluation}

We evaluate the reliability of our evaluation by analyzing whether (1) our evaluation sample represents the broader benchmark population and (2) LLM judges produce consistent measurements.

\subsection{Sample Representativeness}
\label{app:sample}

Evaluating all 1000 DeepResearchGym questions requires excessive compute and LLM token consumption. Due to the resource constraints, we randomly sampled 100 questions and validated this subset's representativeness by comparing its distributional properties against the full dataset, based on the characterizations data from the original Researchy Questions dataset \cite{rosset2025researchy}.
\begin{table*}[t]
\centering
\tiny
\caption{Statistical comparison between the sampled set's 100 questions and the whole DeepResearchGym's 1000 questions. MW represents the Mann-Whitney U test. KS represents the Kolmogorov-Smirnov test. TOST stands for the Equivalence Testing with Two One-Sided Tests. }
\label{tab:statistical_tests}
\resizebox{\textwidth}{!}{%
\begin{tabular}{lcccccccc}
\toprule
\multirow{2}{*}{\textbf{Score Type}} & \textbf{Sampled Set} & \textbf{DeepResearchGym} & \textbf{MW} & \textbf{KS} & \multirow{2}{*}{\textbf{Cohen's d}} & \textbf{TOST $p$-value} & \multirow{2}{*}{\textbf{Equiv?}} \\
& Mean ± SD & Mean ± SD &  $p$-value &  $p$-value & & \textbf{($\delta$=0.5×SD)} & \\
\midrule
Decompositional     & 0.735 ± 0.088 & 0.734 ± 0.087 & 0.9563 & 0.9348 & 0.013  & 0.0000 & Yes \\
Nonfactoid          & 1.022 ± 0.085 & 1.017 ± 0.083 & 0.2800 & 0.5207 & 0.070  & 0.0000 & Yes \\
Ambiguous                  & 0.710 ± 0.697 & 0.797 ± 0.735 & 0.2819 & 0.9898 & -0.119 & 0.0001 & Yes \\
Incompleteness             & 1.390 ± 1.580 & 1.527 ± 1.546 & 0.2330 & 0.9433 & -0.088 & 0.0000 & Yes \\
Assumptive                 & 0.920 ± 2.226 & 0.964 ± 2.122 & 0.1772 & 0.6794 & -0.021 & 0.0000 & Yes \\
Multi-faceted              & 7.140 ± 1.149 & 7.115 ± 1.250 & 0.9458 & 1.0000 & 0.020  & 0.0000 & Yes \\
Knowledge-intensive        & 6.510 ± 1.559 & 6.822 ± 1.307 & 0.0526 & 0.5990 & -0.234 & 0.0057 & Yes \\
Subjective                 & 5.060 ± 2.716 & 4.999 ± 2.666 & 0.6871 & 0.9433 & 0.023  & 0.0000 & Yes \\
Reasoning-intensive        & 6.480 ± 1.144 & 6.600 ± 1.188 & 0.1535 & 0.3537 & -0.101 & 0.0001 & Yes \\
Harmful                    & 0.000 ± 0.000 & 0.000 ± 0.000 & 1.0000 & 1.0000 & ---    & ---    & --- \\
\bottomrule
\end{tabular}%
}
\begin{tablenotes}
\small
\item Note: TOST $p$-value $< 0.05$ indicates two groups are statistically equivalent within margin $\delta$.
\end{tablenotes}
\end{table*}

Table \ref{tab:statistical_tests} shows that Mann-Whitney U and Kolmogorov-Smirnov tests yield $p > 0.05$ across all dimensions, indicating no significant distributional differences. TOST equivalence tests ($p < 0.05$) confirm statistical equivalence within a medium effect size, and Cohen's $d$ values are small (-0.234 to 0.070). These results confirm our sample accurately represents the DeepResearchGym benchmark.

\subsection{LLM Judge Reliability}
\label{app:llm_reliability}

Deep research evaluation is challenging due to report length and complexity, making manual assessment costly. We adopt LLM-as-a-judge protocols and verify their consistency via inter-rater reliability analysis. We ran the judge five times on identical inputs and measured agreement using the Intraclass Correlation Coefficient (ICC).
\begin{table*}[t]
\centering
\caption{Inter-rater reliability analysis using Intraclass Correlation Coefficient (ICC) across three systems under two time constraints. Higher ICC values indicate stronger agreement between multiple runs of the LLM judge.}
\label{tab:icc_scores}
\resizebox{.7\textwidth}{!}{%
\begin{tabular}{llcccc}
\toprule
\textbf{Time Budget} & \textbf{System} & \textbf{Quality} & \textbf{KPR} & \textbf{KPC} & \textbf{Citation Recall} \\
\midrule
\multirow{3}{*}{2 minutes} 
& GPT-Researcher & 0.917** & 0.995** & 0.968** & 0.979** \\
& \method (-AP, -RO) & 0.909** & 0.996** & 0.991** & 0.932** \\
& \method & 0.890* & 0.994** & 0.966** & 0.961** \\
\midrule
\multirow{3}{*}{10 minutes} 
& GPT-Researcher & 0.859* & 0.995** & 0.994** & 0.985** \\
& \method (-AP, -RO) & 0.942** & 0.996** & 0.975** & 0.969** \\
& \method & 0.882* & 0.991** & 0.962** & 0.979** \\
\bottomrule
\end{tabular}%
}
\begin{tablenotes}
\small
\item Note: * Good reliability (0.75 $\leq$ ICC $<$ 0.90), ** Excellent reliability (ICC $\geq$ 0.90).
\end{tablenotes}
\end{table*}

Table \ref{tab:icc_scores} shows ICC scores exceeding 0.85 across all metrics and systems, with most surpassing 0.90 (excellent reliability). Quality metric ICCs range from 0.859 to 0.942, while KPR and KPC exceed 0.96. Citation Recall scores are between 0.932 and 0.985. These high ICC values validate the consistency and reproducibility of our LLM-based evaluation.

\section{Performance Analysis}
\label{app:performance_analysis}

We analyze system performance across time budgets and query characteristics to understand success and failure modes.

\subsection{Time Budget Sensitivity Analysis}
\label{app:sensitivity}

\begin{table*}[t]
\centering
\tiny
\caption{Statistical significance of performance metric changes from 2-minute to 10-minute time budgets, evaluated via independent samples t-tests ($\Delta$ represents mean change).}
\label{tab:time_budget_significance}
\resizebox{\textwidth}{!}{%
\begin{tabular}{lccc|ccc|ccc}
\toprule
\multirow{2}{*}{\textbf{Metric}} & \multicolumn{3}{c|}{\textbf{GPT-Researcher}} & \multicolumn{3}{c|}{\textbf{\ablation}} & \multicolumn{3}{c}{\textbf{\method}} \\
& $\Delta$ & $t$-value & $p$-value & $\Delta$ & $t$-value & $p$-value & $\Delta$ & $t$-value & $p$-value \\
\midrule
\textbf{Quality}\\
Overall & +5.46*** & 9.03 & $<$0.001 & +3.81*** & 5.27 & $<$0.001 & +4.12*** & 7.21 & $<$0.001 \\
Clarity & +4.12*** & 3.49 & $<$0.001 & +1.18 & 1.22 & 0.224 & -0.80 & -0.96 & 0.339 \\
Depth & +1.08** & 2.60 & 0.010 & +0.38 & 0.62 & 0.538 & +0.26 & 1.06 & 0.288 \\
Balance & +1.90*** & 3.93 & $<$0.001 & +2.08** & 3.04 & 0.002 & +0.72 & 1.95 & 0.052 \\
Breadth & +0.00 & 0.00 & 1.000 & +2.52*** & 4.15 & $<$0.001 & +3.32*** & 12.64 & $<$0.001 \\
Support & +18.26*** & 6.90 & $<$0.001 & +16.62*** & 5.93 & $<$0.001 & +18.00*** & 6.45 & $<$0.001 \\
Insightfulness & +7.38*** & 7.10 & $<$0.001 & +0.10 & 0.14 & 0.891 & +3.20*** & 5.67 & $<$0.001 \\
\midrule
\textbf{Relevance}\\
KPR & +9.37*** & 6.40 & $<$0.001 & -0.28 & -0.20 & 0.845 & +3.88** & 2.73 & 0.006 \\
KPC ($\downarrow$) & -0.15 & -0.66 & 0.508 & +0.08 & 0.52 & 0.604 & -0.18 & -1.06 & 0.289 \\
\midrule
\textbf{Faithfulness}\\
Citation Recall & +9.99*** & 10.09 & $<$0.001 & +2.16*** & 3.92 & $<$0.001 & +1.25 & 1.81 & 0.070 \\
\bottomrule
\end{tabular}
}
\begin{tablenotes}
\small
\item Note: Significance levels: *** $p < 0.001$, ** $p < 0.01$, * $p < 0.05$. 
\end{tablenotes}
\end{table*}

We examine performance scaling when increasing the time budget from 2 to 10 minutes (5$\times$). Table \ref{tab:time_budget_significance} shows GPT-Researcher significantly improves across nearly all metrics with more time. \ablation~shows selective gains in quality and breadth but minimal improvement in clarity or depth. \method~displays a distinct pattern: significant overall quality gains (+4.12) are driven by breadth (+3.32), support (+18.00), and insightfulness (+3.20), with minimal changes in clarity and depth. This indicates \method's adaptive planning and orchestration prioritize expanding coverage and strengthening evidence over refining already adequate aspects, effectively optimizing the quality-cost tradeoff.

\subsection{Success and Failure Pattern Analysis}
\label{app:patterns}

\begin{figure*}[t]
    \centering
    \includegraphics[width=\textwidth]{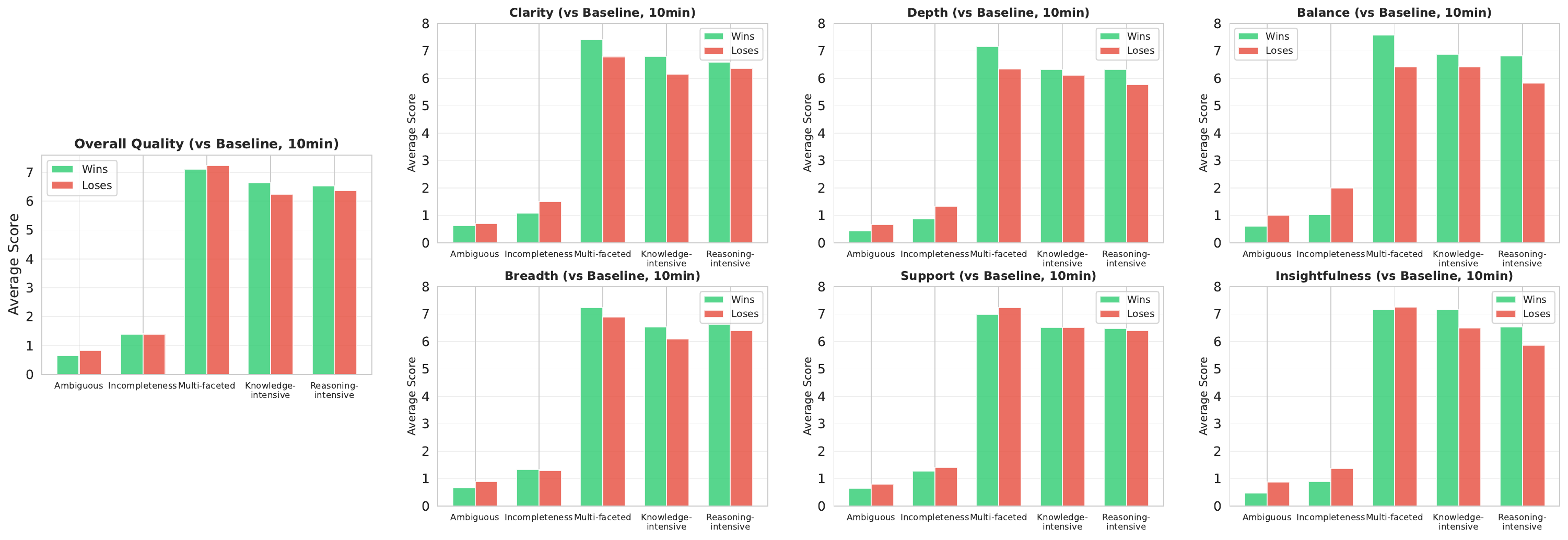}
    \caption{\method performance analysis (v.s. GPT-Researcher baseline, 10 min). We compare the average characterization scores for the 70 questions where \method \textit{Wins} (outperforms the baseline) against the 30 questions where it \textit{Loses}. Scores are plotted across five characterization dimensions (x-axis) from the Researchy Questions dataset \cite{rosset2025researchy} and evaluated with respect to six quality sub-metrics.}
    \label{fig:patterns}
\end{figure*}

We analyze question characteristics to identify where \method~excels or struggles relative to GPT-Researcher.Specifically, we divided the 100 DeepResearchGym questions into two groups based on the 10-minute time budget results: 70 questions where \method~achieved higher overall quality than GPT-Researcher (``Wins"), and 30 questions where it performed worse (``Loses"). For each group, we computed the average scores across five key characterization dimensions from the Researchy Questions dataset: decompositional, multi-faceted, knowledge-intensive, reasoning-intensive, ambiguous, and incomplete. We then analyzed these patterns across six quality sub-metrics: overall quality, clarity, depth, balance, breadth, and support.

Figure \ref{fig:patterns} shows \method~outperforms the baseline on ``knowledge-intensive" and ``reasoning-intensive" questions, achieving better clarity, depth, balance, and breadth. This aligns with our design: adaptive planning broadens exploration, while orchestration focuses on high-value paths. Conversely, \method~underperforms on ``ambiguous" and ``incomplete" questions. Ambiguity may confuse the adaptive planner or cause the orchestrator to prematurely prune paths. This suggests a need to improve orchestration policies for handling ambiguous queries.

\section{Case Analysis}
\label{app:case}
\begin{sidewaysfigure}
    \centering
    \includegraphics[width=\textwidth]{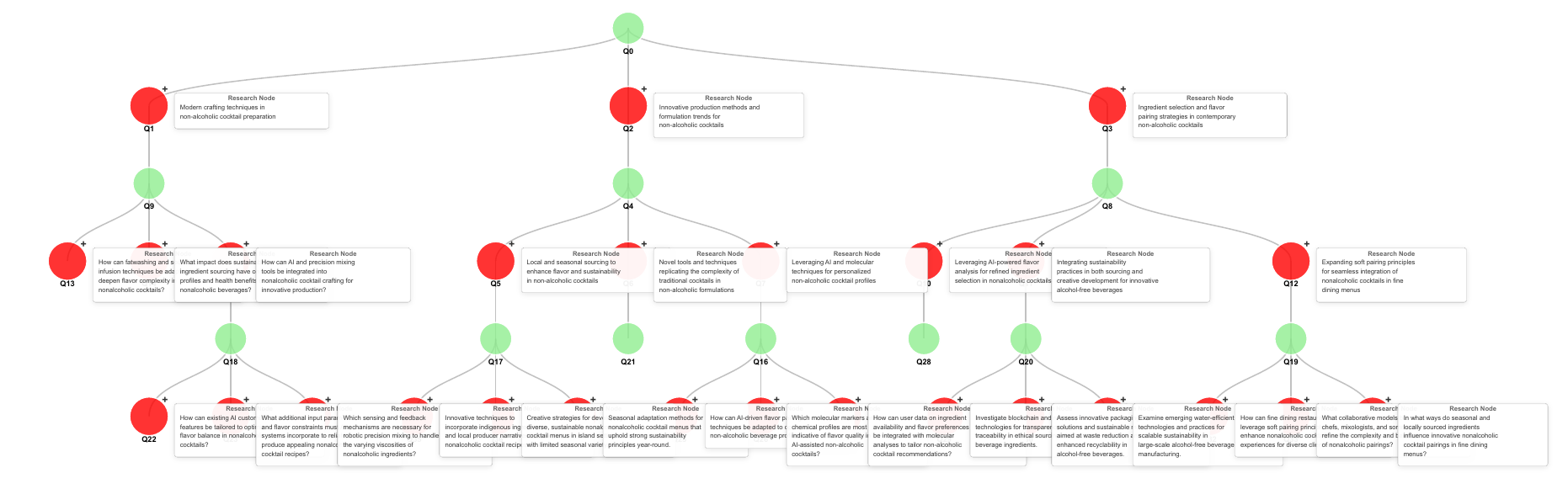}
    \caption{Research tree generated by \method for a broad-topic query: ``\textit{Research Topic: Crafting Techniques for Non-Alcoholic Cocktails. Objective: Investigate current non-alcoholic cocktails to discover innovative production methods and formulations.}''}
    \label{fig:case_1}
\end{sidewaysfigure}

\begin{sidewaysfigure}
    \centering
    \includegraphics[width=\textwidth]{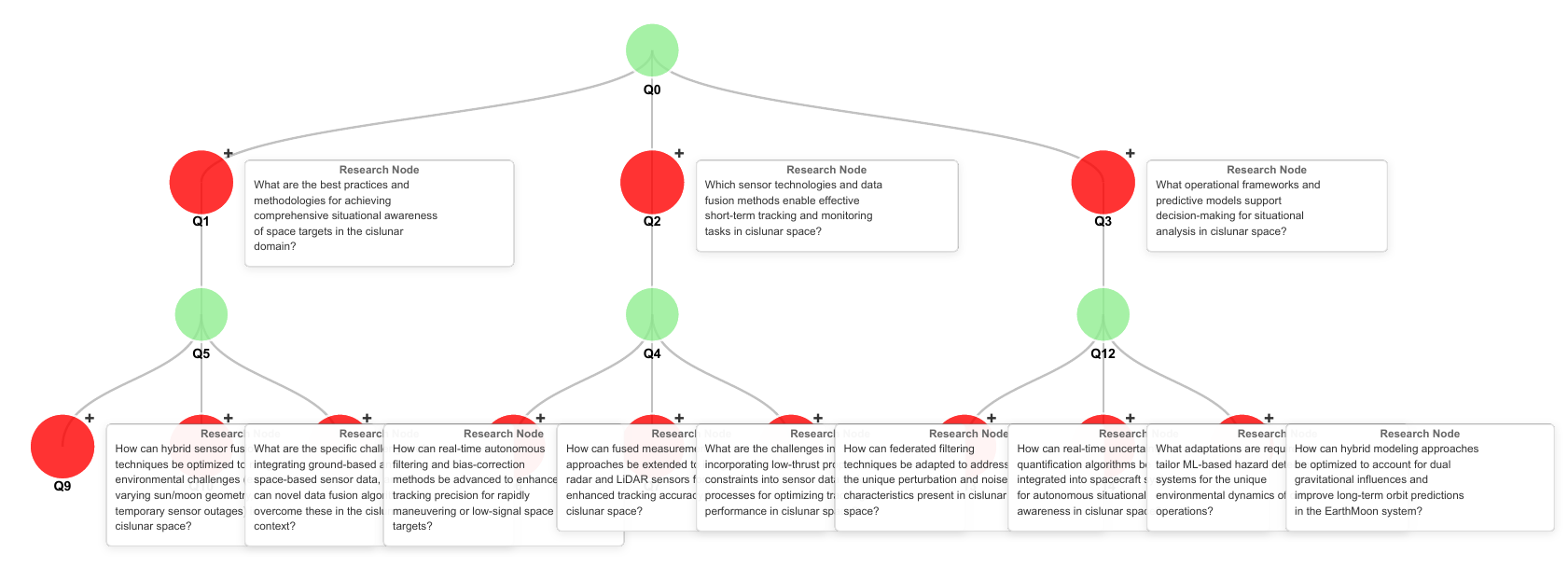}
    \caption{Research tree generated by \method for a narrow, domain-specific query: ``\textit{How to conduct comprehensive and accurate situational awareness of space targets in the cislunar space, and support the effectiveness of short-term cislunar space tracking and monitoring tasks?}''}
    \label{fig:case_2}
\end{sidewaysfigure}

\begin{sidewaysfigure}
    \centering
    \includegraphics[width=\textwidth]{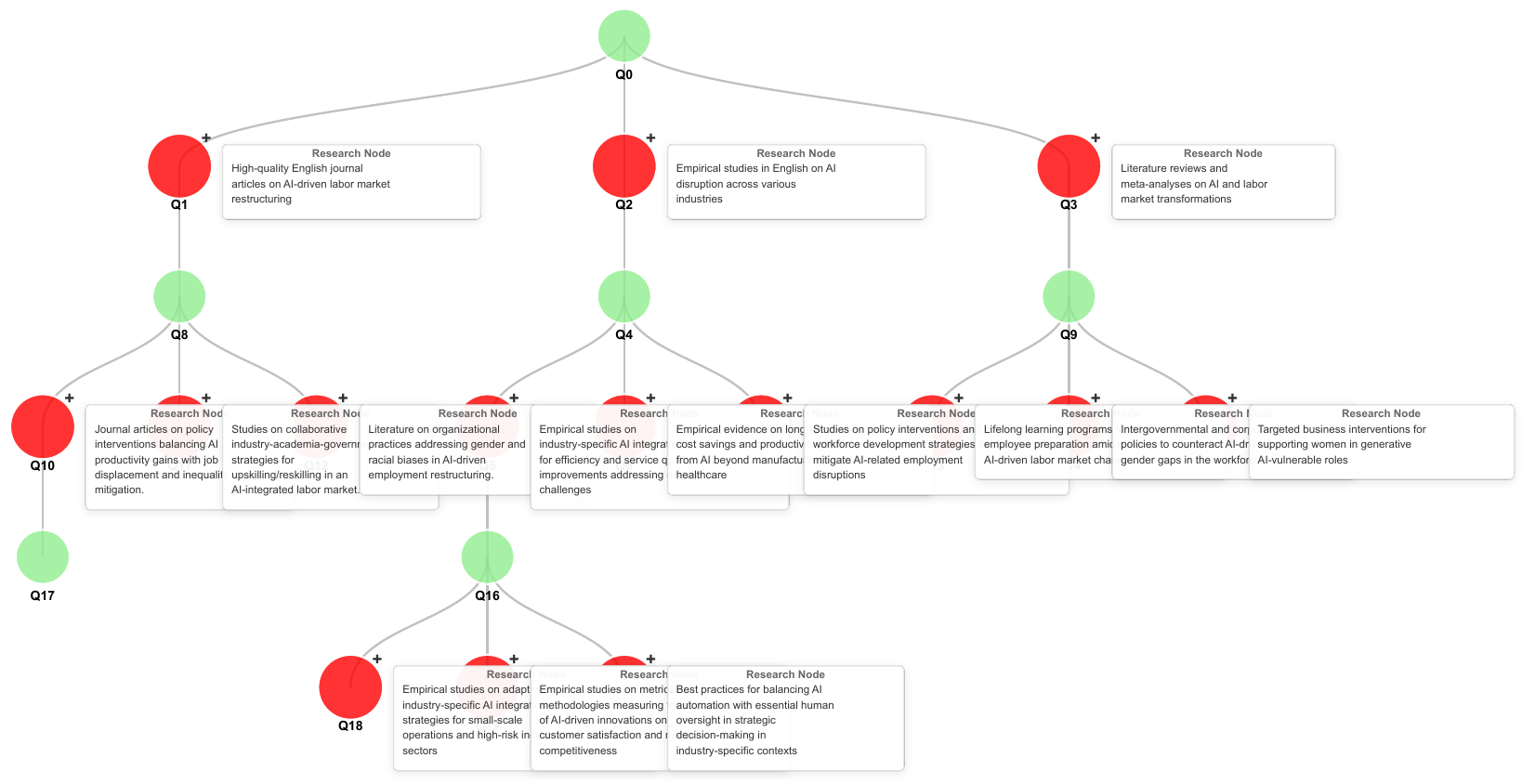}
    \caption{Research tree generated by \method for a user-focused query with explicit demands: ``\textit{Please write a literature review on the restructuring impact of Artificial Intelligence (AI) on the labor market. Focus on how AI, as a key driver of the Fourth Industrial Revolution, is causing significant disruptions and affecting various industries. Ensure the review only cites high-quality, English-language journal articles.}''}
    \label{fig:case_3}
\end{sidewaysfigure}

We illustrate \method's adaptability with three research trees (all 2-minute cutoff). Case 1 (Figure~\ref{fig:case_1}) shows wide expansion for a broad topic (non-alcoholic cocktails), capturing diverse perspectives. Case 2 (Figure~\ref{fig:case_2}) shows a compact tree for a narrow query (cislunar situational awareness), terminating early to save resources. Case 3 (Figure~\ref{fig:case_3}) deepens analysis based on user constraints. \rebuttal{Since thinking models shape planning and orchestration, the framework generates explicit reasoning traces that aid decision interpretation and error tracking.}

In particular, we examined the execution logs and analyzed the reasoning traces to gain a clearer understanding of the decision-making process during runtime orchestration. The user query explicitly requests ``\textit{only cites high-quality, English-language journal articles}". Consequently, when the orchestrator detects that a node does not satisfy this research goal, it terminates the branch early to avoid unnecessary computation. In this specific case, the orchestrator's reasoning trace is shown below.

\begin{example}{}{trace}
The current research progress mainly comprises news articles, reports, and policy discussions rather than high-quality, peer-reviewed English journal articles specifically focused on AI-driven labor market restructuring. While the materials are relevant to the broader topic, they do not meet the stringent criteria of academic journal quality necessary to fully satisfy the research goal. The findings provide useful context and cover significant aspects of AI's impact on the labor market, but they lack the depth, methodological rigor, and academic sourcing expected from high-quality journal articles. Consequently, although there is moderate progress on the subject, key elements of the goal remain unmet, suggesting that it would be more cost-efficient to redirect or deepen the research efforts rather than continuing on the current path.
\end{example}

\section{Generalization to Open-Source Models}
\label{app:opensource}

We evaluate whether \method's benefits transfer to open-source models, which offer cost and privacy advantages but may differ in latency and capability. We test if core mechanisms (adaptive planning, runtime orchestration, and parallelization) remain effective.

\subsection{Experimental Setup with Open-Source Models}
\label{app:opensource_setup}

We evaluated \method~using the open-source Qwen3-235B-A22B-Instruct-2507 (22B activated parameters), deployed locally for all operations. Experimental settings matched our main experiments, with \texttt{gpt-4.1-mini} as the judge, averaged over 5 runs (95\% CI).

\subsection{Results with Open-Source Models}
\label{app:opensource_results}
\begin{table*}[t!]
  \centering
  \setlength{\tabcolsep}{4.0pt}
  \caption{Evaluation of deep research frameworks on DeepResearchGym using the open-source \texttt{Qwen3-235B-A22B-Instruct-2507} model for all the operations. Scores are assessed by \texttt{gpt-4.1-mini-2025-04-14} and averaged over 5 runs. All metrics reported with 95\% confidence intervals.}
  \label{tab:new_results}
  \renewcommand{\arraystretch}{1.15}
  \resizebox{\textwidth}{!}{\begin{tabular}{l c ccccccc cc c}
    \toprule
    & \textbf{Throughput} & \multicolumn{7}{c}{\textbf{Quality}} 
    & \multicolumn{2}{c}{\textbf{Relevance}} 
    & \textbf{Faithfulness} \\ \cmidrule(lr){2-2}
    \cmidrule(lr){3-9} \cmidrule(lr){10-11} \cmidrule(lr){12-12}
     & \# Nodes & \textbf{Overall} & Clarity & Depth & Balance & Breadth & Support & Insight
    & KPR & KPC ($\downarrow$) 
     & Cit. Recall \\
\midrule  
\textbf{\underline{2 minutes}}\\
    Baseline
    & 3.94 ± 0.18
    & 82.26 ± 0.92 & 87.68 ± 1.26 & 91.38 ± 0.39 & 87.54 ± 0.93 & 93.94 ± 0.69 & 46.80 ± 4.20 & 86.20 ± 1.25
    & 65.25 ± 2.09 & \textbf{0.47 ± 0.21}
    & \textbf{79.83 ± 2.02} \\
    \method
    & \textbf{10.90 ± 0.69}
    & \textbf{84.91 ± 0.76} & \textbf{89.22 ± 0.62} & \textbf{96.72 ± 0.41} & \textbf{89.00 ± 0.66} & \textbf{96.22 ± 0.43} & \textbf{48.16 ± 4.04} & \textbf{90.12 ± 0.43}
    & \textbf{69.37 ± 2.00} & 0.93 ± 0.29
    & 74.56 ± 1.65 \\
\midrule
\textbf{\underline{10 minutes}}\\   
    Baseline
    & 15.14 ± 0.31
    & 83.48 ± 0.79 & \textbf{85.56 ± 0.97} & 90.14 ± 0.31 & 87.58 ± 0.67 & 91.04 ± 0.50 & 58.92 ± 3.82 & 87.66 ± 0.98
    & 68.59 ± 1.88 & \textbf{0.47 ± 0.26}
    & \textbf{95.85 ± 0.86} \\
    \method
    & \textbf{68.30 ± 10.04}
    & \textbf{87.68 ± 0.60} & 83.04 ± 0.99 & \textbf{93.84 ± 0.43} & \textbf{88.38 ± 0.57} & \textbf{94.44 ± 0.44} & \textbf{76.64 ± 3.04} & \textbf{89.76 ± 0.29}
    & \textbf{69.99 ± 2.05} & 0.98 ± 0.31
    & 93.78 ± 0.82 \\
    \bottomrule
  \end{tabular}}
\end{table*}

Table \ref{tab:new_results} presents comprehensive results comparing GPT-Researcher baseline and \method~when both use the open-source Qwen3 model. Despite the higher latency of locally-served Qwen3, \method~achieves substantial throughput gains: 2.77× improvement (10.90 vs 3.94 nodes) under 2-minute budget and 4.51× improvement (68.30 vs 15.14 nodes) under 10-minute budget. These gains demonstrate that \method's parallel orchestration mechanisms effectively compensate for individual operation latency. 

Most remarkably, \method's 2-minute execution achieves higher overall quality (84.91 ± 0.76) than the baseline's 10-minute execution (83.48 ± 0.79), representing a 5× speedup while maintaining superior quality. This finding closely mirrors our results with proprietary models (Table 1), confirming that \method's architectural benefits are model-agnostic.

Examining individual metrics reveals interesting patterns. The 2-minute executions achieve higher scores in clarity, depth, and breadth compared to 10-minute executions. This counterintuitive result may reflect limitations in the open-source model's ability to effectively manage long contexts during information aggregation. However, 10-minute executions significantly outperform in support (+28.48 points) and citation recall (+19.22 points), indicating that longer execution still substantially benefits evidentiary quality and faithfulness.

Overall, these results validate that \method can successfully transfer to open-source models, making the approach practical for diverse deployment scenarios.

\section{Economic Cost Analysis}
\label{app:cost}

While \method~introduces additional computational overhead through runtime orchestration (periodic quality and goal satisfaction assessments), it also prunes low-value research paths early, potentially reducing overall costs. Understanding the net economic impact requires detailed analysis of token usage across all components. This section quantifies the monetary costs associated with different systems and examines whether orchestration overhead is offset by efficiency gains.
\begin{table*}[t!]
  \centering
  \setlength{\tabcolsep}{3.5pt}
  \caption{LLM token usage and API cost analysis on DeepResearchGym with different time budgets. All values are averaged over 100 questions. Orchestration metrics only apply to \method.}
  \label{tab:token_usage}
  \renewcommand{\arraystretch}{1.15}
  \resizebox{.8\textwidth}{!}{\begin{tabular}{l ccc ccc}
    \toprule
    & \multicolumn{3}{c}{\textbf{Orchestration Usage}} & \multicolumn{3}{c}{\textbf{Total Usage}} \\
    \cmidrule(lr){2-4} \cmidrule(lr){5-7}
    & \textbf{Input Tokens} & \textbf{Output Tokens} & \textbf{Cost (\$)}
    & \textbf{Input Tokens} & \textbf{Output Tokens} & \textbf{Total Cost (\$)} \\
    \midrule  
    \textbf{\underline{2 minutes}} \\
    GPT-Researcher
    & --
    & --
    & --
    & 97,170
    & 5,428
    & 0.1258 \\
    \ablation
    & --
    & --
    & --
    & 176,147
    & 9,355
    & 0.2218 \\
    \method
    & 9,632
    & 1,788
    & 0.0185
    & 155,409
    & 10,031
    & 0.2071 \\
    \midrule
    \textbf{\underline{10 minutes}} \\   
    GPT-Researcher
    & --
    & --
    & --
    & 299,093
    & 18,432
    & 0.3941 \\
    \ablation
    & --
    & --
    & --
    & 424,488
    & 28,496
    & 0.5629 \\
    \method
    & 25,646
    & 4,756
    & 0.0491
    & 487,023
    & 36,409
    & 0.6811 \\
    \bottomrule
  \end{tabular}}
\end{table*}

Table \ref{tab:token_usage} provides a detailed breakdown of LLM API usage and costs averaged across 100 DeepResearchGym questions. We report both orchestration-specific costs (only applicable to \method) and total system costs including all operations (planning, web search, retrieval, reasoning, and report generation).

Under the 2-minute budget, \method's orchestration layer consumes 9,632 input tokens and 1,788 output tokens per question on average, costing \$0.0185. This represents only 8.9\% of the total system cost (\$0.2071). Under the 10-minute budget, orchestration costs rise to \$0.0491 but remain just 7.2\% of total costs (\$0.6811). These figures demonstrate that runtime orchestration introduces minimal overhead relative to overall system costs.

Comparing total costs across systems reveals important tradeoffs. GPT-Researcher baseline costs \$0.1258 (2-min) and \$0.3941 (10-min). The \ablation~system (without adaptive planning and orchestration) costs \$0.2218 (2-min) and \$0.5629 (10-min), showing that adding parallelization alone increases costs by 76\% and 43\% respectively. \method~costs \$0.2071 (2-min) and \$0.6811 (10-min). While \method~incurs 65\% and 73\% higher costs than the baseline, it explores 2.1× and 4.5× more research nodes while achieving superior quality (Table 1). Importantly, \method's 2-minute execution (\$0.2071, Quality: 83.82) achieves comparable quality to the baseline's 10-minute execution (\$0.3941, Quality: 83.74) at roughly half the cost, demonstrating substantial economic efficiency.

The key insight is that \method~achieves better cost-quality tradeoffs by intelligently allocating computational resources. While absolute costs increase moderately, the system delivers disproportionate quality improvements by focusing resources on high-value research paths and pruning redundant exploration. The orchestration overhead (7-9\% of costs) is far outweighed by the gains from improved resource allocation.

\twocolumn

\onecolumn

\end{document}